\documentclass[useAMS,usenatbib]{mn2e}

\usepackage{graphicx,mathptm}
\usepackage{times}
\usepackage{natbib}

\newcommand{\lesssim}{\la} 
 
\newcommand{\lognlogs}{log$N$-log$S$}
\newcommand{\sbunits}{erg s$^{-1}$ cm$^{-2}$ deg$^{-2}$}
\newcommand{\lumunits}{erg s$^{-1}$ cm$^{-2}$}

\def\aap{A\&A}
\def\apj{ApJ}

\def\apjl{ApJ}
\def\mnras{MNRAS}
\def\araa{ARA\&A}
\def\aj{AJ}

\def\nat{Nat}

\def\apjs{ApJS}
\def\jcap{JCAP}        

\title[]
{Properties of the diffuse X-ray background in a 
high-resolution hydrodynamical simulation}
\author[M. Roncarelli et al.]
{M. Roncarelli$^1$, 
L. Moscardini$^{1}$, 
P. Tozzi$^{2}$,
S. Borgani$^{3,4}$,
L.M. Cheng$^{5}$,
A. Diaferio$^6$,
\newauthor
K. Dolag$^{7}$,
G. Murante$^8$
 \\~\\ 
$^1$ Dipartimento di Astronomia, Universit\`a di Bologna,
via Ranzani 1, I-40127 Bologna, Italy
(mauro.roncarelli,lauro.moscardini@unibo.it)\\
$^2$ INAF, Osservatorio Astronomico di Trieste, via Tiepolo 11,
  I-34131 Trieste, Italy (tozzi@ts.astro.it)\\
$^3$ Dipartimento di Astronomia, Universit\`a di Trieste, via
  Tiepolo 11, I-34131 Trieste, Italy (borgani@ts.astro.it)\\
$^4$ INFN -- National Institute for Nuclear Physics, Trieste,
  Italy\\ 
$^5$ Institute of Theoretical Physics, Chinese Academy of Sciences,  
  Beijing 100080, China (clm@itp.ac.cn)\\
$^6$ Dipartimento di Fisica Generale ``Amedeo Avogadro'', Universit\`a
  di Torino, via Giuria 1, I-10125 Torino, Italy (diaferio@ph.unito.it) \\
$^7$ Max-Planck Institut fuer Astrophysik,
Karl-Scwarzschild Strasse 1, D-85748 Garching, Germany
(kdolag@mpa-garching.mpg.de)\\
$^8$ INAF, Osservatorio Astronomico di Torino, Strada Osservatorio 20,
  I-10025 Pino Torinese, Italy (murante@to.astro.it)\\
}
\begin{document}


\pagerange{\pageref{firstpage}--\pageref{lastpage}} \pubyear{2005}

\maketitle

\label{firstpage}

\begin{abstract}
 
We study the properties of the diffuse X-ray background by using the
results of a cosmological hydrodynamical simulation of the concordance
$\Lambda$CDM model. The simulation follows gravitational and gas dynamics 
and includes a treatment of physical processes like
radiative cooling, star formation and supernova feedback. From the simulation
outputs, we produce a set of two-dimensional maps of the intergalactic medium 
X-ray emission integrated over redshift. We find that the signal in
the soft (0.5-2 keV) band is lognormally distributed with a mean
intensity of about $4 \times 10^{-12}$ \sbunits; approximately 40 per
cent of the emission originates from warm-hot gas (defined as baryons with 
$10^{5}<T<10^{7} $ K), 
and 90 per cent comes from structures at $z<0.9$.  Since the
spectrum is soft, being mostly provided by the intergalactic medium at low
temperature, the total mean intensity in the hard (2-10 keV) X-ray
band is smaller by a factor of about 4. In order to constrain 
the physical processes included in our simulation, we compare our results with the
observed upper limit $(1.2\pm 0.3) \times 10^{-12}$ \sbunits \ 
of the soft X-ray emission due to diffuse gas. To
this purpose, we remove the contributions of observable extended
objects (groups and clusters of galaxies) from the simulated maps by
adopting different detectability criteria which are calibrated on
the properties of systems at intermediate redshifts observed by \emph{Chandra}. 
We show that the simulated diffuse soft X-ray emission
is consistent with the present observed upper limit. However,
if future measurements will decrease the level
of the unresolved X-ray background by a factor of two, a more efficient feedback
mechanism should be required to suppress the soft emission of the gas
residing within filaments and group-size haloes.
\end{abstract}



\section{Introduction} \label{sect:intro}

A general result of hydrodynamic simulations of cosmic structure
formation is that, during the formation of galaxy clusters via
gravitational collapse, a large amount of gas remains out of these
structures at the present time. An important fraction of the total baryons
of the universe (roughly 40-50 per cent) undergoes a process of
shock-heating that starts at $z \sim 2$ and heats the baryons to an
intermediate temperature of $10^5$ to $10^7$ K: this gas phase is often
referred to as the Warm-Hot Intergalactic Medium (WHIM) \citep[see, e.g.,
][]{dave2001,croft2001}. Simulations also show that these baryons are
not uniformly distributed: they constitute a filamentary network
linking the largest virialized objects, the so-called cosmic web
\citep{bond1996}. The existence of these structures is believed 
to be the solution to the missing baryon problem \citep[see, e.g.,
][]{cen1999}, namely the fact that more than half the normal matter is yet to be
detected by instruments: in fact, the density of filaments is expected to
be very low and consequently their detection is made difficult by both their
extremely low surface brightness and projection effects \citep[see,
however, some first claimed detections in X-ray analyses:
][]{zappacosta2002,markevitch2003,finoguenov2003}. In any case, the
intermediate temperature of the cosmic web gas suggests that it can
produce a non-negligible X-ray emission through thermal
bremsstrahlung.  Given the physical state of the gas, this emission is
expected to appear mostly as a diffuse background at soft
energies.

It is now generally accepted that some of the observed cluster X-ray
properties, like the scaling relations between mass, X-ray luminosity
and temperature and their redshift evolution, the temperature
profiles, or the entropy excess in the central regions of poor
clusters and groups, cannot be reproduced by a simple model of
gasdynamics based on gravitational heating alone \citep[see the discussion
in ][ and references therein]{rosati2002,voit2005}. In an attempt to
explain these discrepancies, new processes have been included in
the models of cluster formation, largely increasing the complexity of
the physics of the intracluster medium (ICM): radiative cooling, energy
feedback from supernovae (SN\ae ) and active galactic nuclei (AGN),
thermal conduction, turbulence, magnetic fields, etc. \citep[see,
e.g.,][] {bower1997,cavaliere1998,balogh1999,pearce2000,
bryan2000,tozzi2001,muanwong2001,voit2001b,babul2002,voit2002,
tornatore2003,jubelgas2004,dolag2005,dimatteo2005}.  All these
processes affect the properties of both the gas within virialized objects, like galaxy groups
and clusters, and the diffuse gas component which has low density and
temperature.  Therefore, the expected X-ray emission from the
cosmic web can sensitively change when the parameters describing the
ICM physics are varied.  For these reasons, observational data on the
soft X-ray background emitted by diffuse gas can be used as a probe
of these physical processes, as suggested by different authors
\citep[see, e.g.,][]{voit2001a,bryan2001,xue2003}.

Thanks to the latest generation of X-ray satellites (\emph{Chandra} and
\emph{XMM-Newton}),  the observational picture about the cosmic X-ray
background (hereafter XRB) in the soft (0.5-2 keV) energy band has
become much clearer.  Very deep pointed observations, like the 1-2 Ms
\emph{Chandra} Deep Fields (CDFs) \citep{giacconi2002,alexander2003}
and the Lockman Hole \emph{XMM-Newton} data \citep{worsley2004},
showed that the soft XRB is largely produced by individual discrete
sources, mostly AGN. Recent estimates of the
percentage of the resolved contribution range between 85 and 95 per cent
\citep[see,  e.g.,][]{bauer2004,worsley2005}.
These results give a conservative estimate of approximately 
$(1.2\pm 0.3) \times 10^{-12}$ \sbunits \ for the intensity of the still 
unresolved soft X-ray background 
(see the discussion in Section \ref{sect:obs_est}): therefore this value 
can be used to set a stringent upper limit to the background from the 
diffuse IGM.

The availability of these new observational constraints suggests to
re-consider the problem of the soft X-ray emission from the cosmic web
and its implications for the ICM physics by updating an 
analysis made by \cite{croft2001}. The main goal of our paper is to
check the consistency of our model with the observed limits by using 
the outputs of a cosmological hydrodynamic simulation
\citep{borgani2004}, which, besides gravitational and gas dynamics, 
includes a treatment of the ICM with radiative cooling,
star formation and SN feedback.  Our previous analyses
\citep{borgani2004,ettori2004,diaferio2005} showed that the results of
this simulation are in encouraging agreement with some of the most
significant observed properties of clusters, such as the
mass-temperature and the X-ray luminosity-temperature relations;
however, the feedback mechanism does not appear to be efficient enough to avoid
the production of overluminous groups and poor clusters.

The paper is organized as follows. In Section \ref{sect:mapsim} we
describe the characteristics of our simulation 
and present the procedure used to create the set of
X-ray maps. The general statistical properties of
the XRB extracted from these maps are discussed in Sections
\ref{sect:softX} and \ref{sect:hardX} for the soft (0.5-2 keV) and
hard (2-10 keV) X-ray bands, respectively. Section \ref{sect:comparison} is 
devoted to the comparison of our results with previous works. In Section
\ref{sect:diffusegas} we review the observational estimates of the
contribution of diffuse gas to the soft XRB and we compare them with the
results of our mock maps in order to obtain qualitative constraints
on the modeling of the physical processes included in our
hydrodynamical simulation.  We conclude in 
Section \ref{sect:conclusions}.


\section{The map construction} \label{sect:mapsim}

\subsection{The hydrodynamical simulation}

The study of the properties of the diffuse X-ray emission from the
large-scale structure of the universe and its comparison with 
observations requires the analysis of the total
emission integrated over redshift.  Therefore, we need to simulate
the entire volume of the past light-cone seen by an observer located at
$z=0$.  We construct the light-cone, which enables
the production of simulated maps of the X-ray intensity, by
using the outputs of a cosmological hydrodynamical simulation.
Specifically, we use the results of the simulation by 
\cite{borgani2004}, which considers the concordance cosmological model,
i.e a flat $\Lambda$CDM model dominated by the presence of the
cosmological constant ($\Omega_{\rm m}=0.3$, $\Omega_{\Lambda}=0.7$),
with a Hubble constant $H_0=h\ 100$ km s$^{-1}$Mpc$^{-1}$ and
$h=0.7$, and a baryon density $\Omega_{\rm b}=0.04$. The initial
conditions were set by the cold dark matter (CDM) power spectrum and
were normalized by assuming $\sigma_8=0.8$.  The run was carried
out with the TreeSPH code GADGET-2 \citep{springel2001,springel2005}
and followed the evolution of $480^{3}$ dark matter
(DM) particles and as many gas particles from $z=49$ to $z=0$.
The cubic box is 
$192 h^{-1}$ Mpc on a side, and the DM and gas particles have mass $m_{\rm DM}=
4.62 \times 10^9 h^{-1} M_\odot$ and $m_{\rm gas}= 6.93 \times 10^8
h^{-1} M_\odot$, respectively. The
Plummer-equivalent gravitational softening is $\epsilon$=7.5
$h^{-1}$ kpc at $z=0$, fixed in physical units between $z=2$ and $z=0$, and 
fixed in comoving units at earlier times.

The simulation, besides gravity and non-radiative hydrodynamics,
includes a treatment of the processes that influence the physics of
the ICM: star formation, by adopting a
sub-resolution multiphase model for the interstellar medium
\citep{springel2003}, feedback from SN\ae\ with the effect of weak
galactic outflows, radiative gas cooling and heating by a uniform,
time-dependent, photoionizing UV background.  This run produced one
hundred outputs, equally spaced in the logarithm of the expansion factor,
between $z=9$ and $z=0$.

\subsection{The map-making procedure}

In order to create mock maps of the X-ray emission of the structures
within the past light-cone, we follow the same technique
adopted by \cite{croft2001} \citep[see also similar applications for
maps of the Sunyaev-Zel'dovich effect,
e.g.][]{dasilva2001,springel2001,white2002}. The method is based on
the replication of the original box volume along the line of sight.
Since we assume a flat cosmological model, we use the comoving
coordinates for the projection because the light trajectory is a
straight line in this reference frame.  We build past
light-cones which extend to $z=6$. Even if this maximum redshift can appear
too large to still yield a significant signal in the X-ray band (this will
be confirmed by our analysis in Section \ref{sect:softX}), this choice
is done to obtain, at the same time, maps for the thermal and
kinetic Sunyaev-Zel'dovich effects (not discussed in this paper), for
which there is a non-negligible contribution from the high-redshift gas.
The extension of the light-cone corresponds to a comoving distance of
approximately $5,770\ h^{-1}$ Mpc, so we need to stack 
the simulation volume roughly 30 times. However,
in order to obtain a better redshift sampling,
rather than stacking individual boxes, we adopt the following
procedure.  We divide the simulated box at each output redshift into three equal
slices along the line of sight (each of them with a depth of $64\
h^{-1}$ Mpc); for the stacking procedure, we choose 
the slice extracted from the simulation output 
that better matches the redshift of the central point of
the slice. Our light-cones are thus built with 91 slices extracted from 82 different
snapshots.

The necessity of avoiding the repetition of the same structures along
the line of sight requires a randomization of the boxes used to build
our maps: since our simulation assumes periodic boundary conditions,
for each box entering the light-cone we combine a process of random
recentering of the coordinates with a 50 per cent probability of
reflecting each axis. The slices belonging to the same box undergo the
same randomization process to avoid spatial discontinuities between
them: this allows to retain the entire information on the structures
within the box and strongly reduces the loss of power on larger
scales.  By varying the initial random seeds we also 
obtain different light-cone reconstructions: we use this technique to
produce ten different realizations that we use to assess the
statistical robustness of our results.

\begin{figure}
\includegraphics[width=0.45\textwidth]{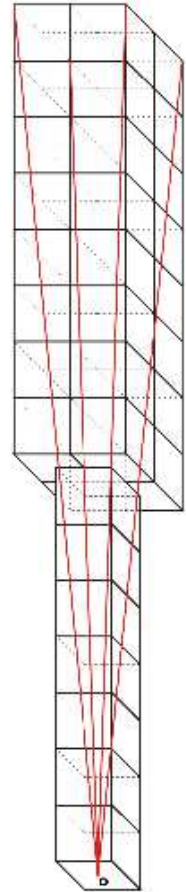}
\caption{
Sketch of the configuration adopted to realize the light-cone.  The
observer is located at the position $O$ at the centre of the lowest
side of the first box.  The past light-cone is obtained by stacking the
comoving volumes of the simulation outputs at the corresponding
redshift.  In order to obtain a large field of view of size $3.78^2$
deg$^2$, starting at $z=1.4$ we use four replications of the
box at the same redshift.  The red lines show the volume in the light 
cone corresponding to the field of view.  }
\label{fig:cubes}
\end{figure}

In order to have maps covering a larger field of view, we 
replicate the boxes four times across the line of sight starting at
comoving distances larger than half the light-cone extension
(i.e. larger than about $2,900\ h^{-1}$ Mpc, corresponding to
$z>1.4$).  The strategy to pile up boxes is shown in
Fig. \ref{fig:cubes}.  In this way we obtain maps 
3.78 deg on a side.  Every map contains 8,192 pixel on a side:
consequently our resolution is 1.66\arcsec, roughly three times the
resolution of the CDFs at the aimpoint.

\begin{figure*}
\includegraphics[width=0.45\textwidth]{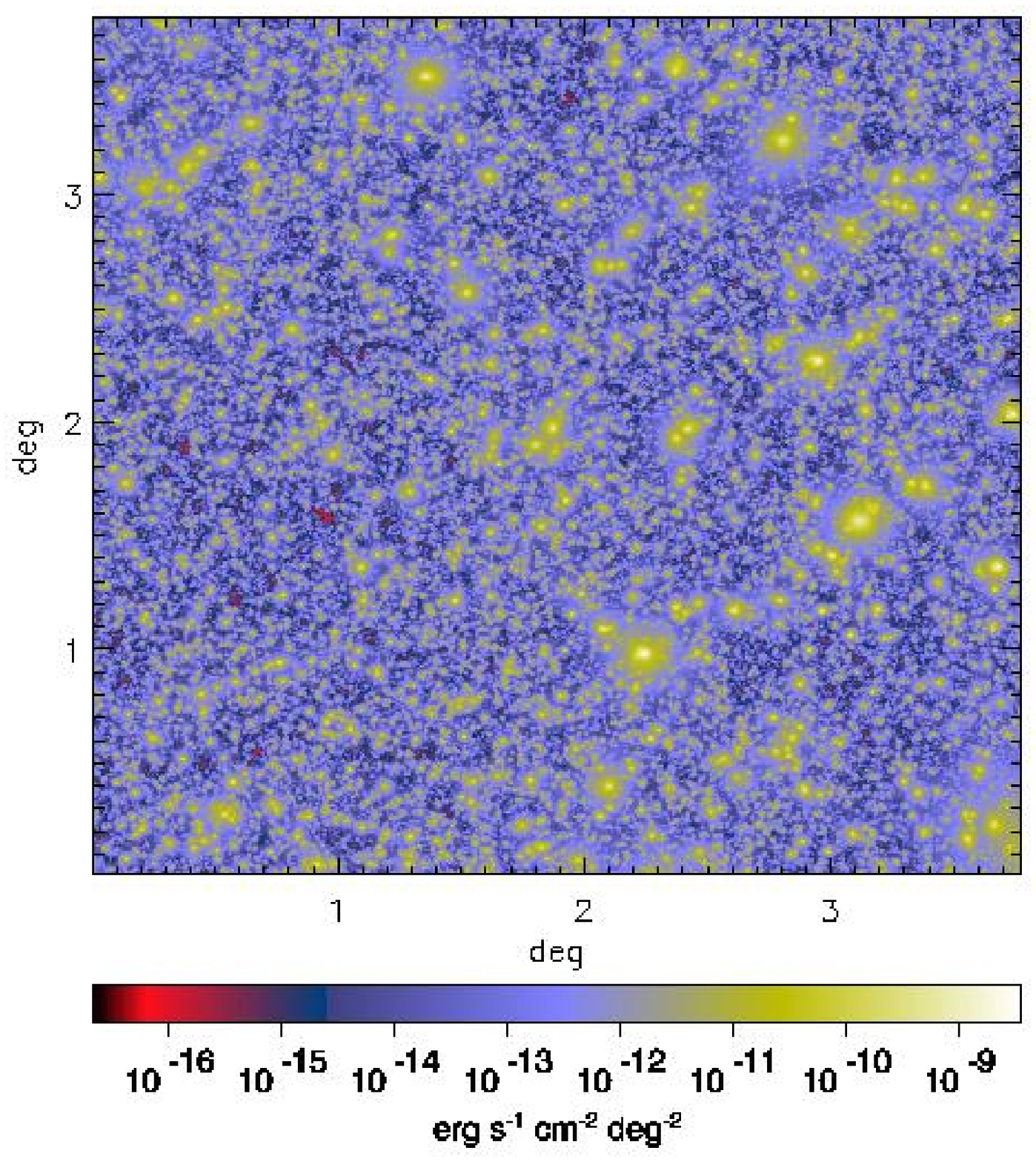}
\includegraphics[width=0.45\textwidth]{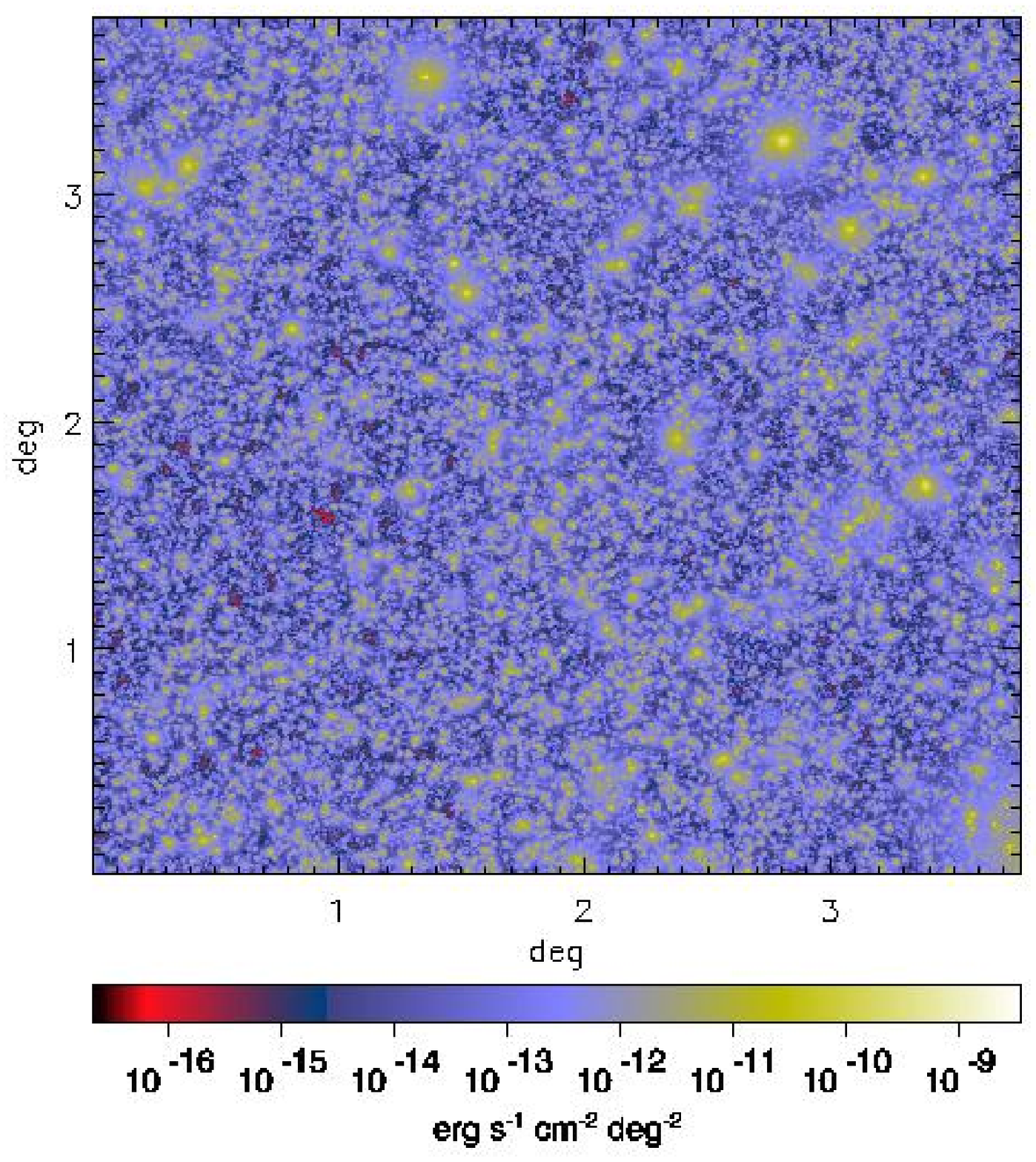} 
\caption{
Maps of the soft (0.5-2 keV) X-ray intensity obtained by considering
all the gas particles (IGM, left panel), and only the gas particles with 
temperature in the range $10^{5}<T<10^{7}$ K (WHIM, right panel).  The
maps are 3.78 deg on a side and the pixel size is
$(1.66\arcsec)^2$. These two maps refer to the same realization of the past
light-cone.  }
\label{fig:softmaps}
\end{figure*}

We now need to calculate the contribution, to the X-ray emission, of every
gas particle that lies within the light-cone volume. The X-ray
luminosity of the $i$-th particle in a given energy band $[E_1,E_2]$,
as measured at $z=0$, is calculated as
\begin{equation}
L_{X,i} = (\mu m_p)^{-2} x_e m_i \rho_i \Lambda(T_i,Z_i,E_1',E_2')\ ,
\label{eq:luminosity}
\end{equation}
where $\mu$ is the mean molecular weight in units of the proton mass
$m_p$, $x_e$ represents the ratio between the number density of free
electrons and hydrogen nuclei ($n_e/n_H$), $m_i$ and $\rho_i$ are the
particle mass and density, respectively. Since only the gas with
$T>10^5$ K gives a significant contribution to the X-ray emission, we
can safely assume full ionization of hydrogen and helium for all the
particles: therefore the values of $\mu$ and $x_e$ depend only on the
metallicity (notice that $\mu=0.588$ and $x_e=1.158$ for zero
metallicity). The cooling function $\Lambda$ is computed by using the
plasma emission model by \cite{raymond1977} as a function of
temperature $T_i$ and metallicity $Z_i$ of the particle and
depends on the energy interval $[E_1',E_2']$; this interval is 
$K$-corrected for redshift: $E_{1,2}'=E_{1,2}(1+z)$.

As we will discuss in the following section, a sizeable contribution
to the soft X-ray background comes from relatively warm particles,
with temperature below $10^7$ K. At these temperatures, the emissivity
from metal lines becomes non negligible and, therefore, one needs to
account for their contribution. In its original implementation, GADGET-2
includes a prescription to generate metals from SN explosions. This
model assumes that only SN-II contribute to the chemical enrichment,
under the assumption of instantaneous recycling, i.e. metals are
released instantaneously when new stars form and the
effect of stellar life-times is neglected \citep[see, e.g., ][ for a more detailed
implementation of chemical enrichment in GADGET-2]{tornatore2004}. In
Eq.  (\ref{eq:luminosity}) we adopt the value of metallicity
$Z_i$ yielded by this prescription to compute the luminosity of each gas particle.

We then calculate the contribution of the $i$-th gas particle to the
X-ray intensity $I_{X,i}$ as
\begin{equation}
I_{X,i}=L_{X,i}/(4\pi d_L(z)^2 A) \ ,
\end{equation}
where $d_L(z)$ is the luminosity distance and $A$ is the angular area
covered by a pixel. This quantity is then distributed over the pixels
by using an SPH smoothing kernel given by
\begin{equation}
W(x)\propto \left\{ \begin{array}{ll}
1-6x^2+6x^3, & 0\le x<0.5 \nonumber \\
2(1-x)^3, & 0.5<x\le1 \\
0, & x>1. \nonumber
\end{array} \right.
\end{equation}
In the previous expressions 
$x\equiv\Delta\theta/\alpha_i$, where $\Delta\theta$ is the angular
distance between the pixel centre and the projected particle position
and $\alpha_i$ is the angle subtended by the particle smoothing length
provided by the hydrodynamical code.  In order to conserve the
total intensity emitted by each particle we normalize to unity the sum 
of the weights $W$ over all ``touched'' pixels.  Finally,
the intensity for a given pixel is obtained by summing over all the
particles inside the light-cone. 

To avoid spurious effects in the computation of the X-ray
intensity, we exclude the particles having a mean
electron density $n_e> 0.26\,h^2$cm$^{-3}$. According to the model of 
\cite{springel2003}, these particles are assumed to be composed by a
hot ionized phase and a cold neutral phase, whose relative amounts
depend on the local temperature and density. Since these particles are
meant to account for the multi-phase nature of the interstellar medium,
we correctly exclude them from the computation of the X-ray emissivity.
 

\section{The properties of the soft X-ray background}\label{sect:softX}

Examples of intensity maps obtained by adopting the method described
above are displayed in Fig. \ref{fig:softmaps} for the emission in the
soft (0.5-2 keV) band.  Following \cite{croft2001}, we distinguish two
X-ray contributions: the contribution from the intergalactic medium (IGM, left panel),
i.e. from all the gas particles, and the contribution from the warm-hot intergalactic
medium (WHIM, right panel), i.e. from the gas particles having a
temperature between $10^{5}$ and $10^{7}$ K. The IGM map is dominated
by several extended bright galaxy clusters (with flux in the range
$10^{-12}-10^{-11}$ \lumunits) that give significant contributions to
the total flux; there is also a large number of smaller
structures, corresponding to nearby galaxy groups or distant fainter
clusters. These smaller objects give the dominating contribution 
in the WHIM maps, because the WHIM does not contain the hottest particles;
on the contrary, in these maps, the
brightest galaxy clusters are much less prominent. Note that,
because of projection effects, neither map shows evident signatures of
emission from filamentary structures.

\begin{figure}
\includegraphics[width=0.45\textwidth]{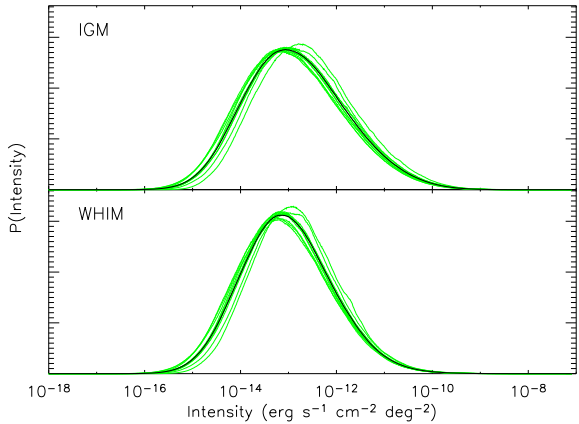}
\caption{
Distribution of pixel values of the intensity in the soft X-ray band.
Upper and lower panels refer to the IGM and the WHIM, respectively. The thin
lines show the results of ten different realizations; the thick
solid line is the corresponding average.  }
\label{fig:softdistr}
\end{figure}

The distribution of the values assumed by the intensity (considering
the 1.66\arcsec pixels) is shown in Fig. \ref{fig:softdistr} for both 
the IGM and the WHIM. The thin lines refer to each of the ten
light-cone realizations, while the thick solid line shows their
average.  A small dispersion between the different realization is
evident: this is mainly due to the inclusion, in the maps, of a larger
or smaller number of galaxy clusters at relatively low redshifts.  In
general the two averaged distributions for ${\cal I}\equiv \log I_X$
(\sbunits) are very close to a Gaussian distribution. In particular we
find $\overline{\cal I}=-12.81$ with a corresponding r.m.s. of 1.05
for the IGM, and $\overline{\cal I}=-13.01$ with a corresponding
r.m.s. of 0.94 for the WHIM; the skewness is 0.30 and 0.25, for the
IGM and the WHIM respectively.  Of course the distributions differ for
larger fluxes: this is due to the fact that the flux of the brightest
structures comes from gas at high temperature.

The values of the intensity averaged over the ten map realizations is
reported in Table \ref{tab:meanave} for both the IGM and the WHIM. The quoted
errors are the dispersions computed in fields of 1 deg$^2$ to allow a
direct comparison with the corresponding results reported in
\cite{croft2001}.  We find that the mean intensity for the IGM is about $4
\times 10^{-12}$ \sbunits, a factor 1.8 larger than in
\cite{croft2001}. The corresponding dispersion is also larger: about
50 per cent compared to about 20 per cent in \cite{croft2001}.  Again, we checked 
that the
high map-to-map spread is due to the presence of some bright galaxy
clusters very close to the observer and giving a strong contribution
to the total emission: this effect is also evident in the pixel
distributions of individual maps shown in Fig. \ref{fig:softdistr}.
By looking at the WHIM contribution, we find that its mean intensity is
about 40 per cent of the total bremsstrahlung emission in the soft
band (see Table
\ref{tab:meanave}). Again this value is larger
than in the analysis of \cite{croft2001}, who reported a value of $4.15
\times 10^{-13}$ \sbunits, corresponding to about 20 per cent of the
total intensity of the IGM.  However, we emphasize that we 
cannot make a direct comparison 
between these results and those presented in \cite{croft2001}
because the simulation analyzed in that paper
considers a much smaller box ($50 h^{-1}$ Mpc on a side), assumes a
slightly different cosmological model (a flat universe with
$\Omega_{\rm m}=0.4$ and a primordial spectral index $n=0.95$) and,
most importantly, the included physical processes are not the same 
(see the discussion in \ref{sect:comparison}).

\begin{table}
\begin{center}
\caption{
Average value of the intensity $I_X$ in the two different X-ray bands
for the IGM and the WHIM. The average is computed over ten different map
realizations; the quoted errors are the r.m.s.  in fields of 1
deg$^2$. }
\begin{tabular}{lcc}
\hline
\hline
Energy band & IGM (total) & WHIM \\
 & (\sbunits) & (\sbunits) \\
\hline

{\scshape Soft} (0.5 - 2 keV) & $(4.06 \pm 2.19) \times 10^{-12}$ & 
$(1.68 \pm 0.62) \times 10^{-12}$ \\

{\scshape Hard} (2 - 10 keV) & $(1.01 \pm 1.53) \times 10^{-12}$ & 
$(2.92 \pm 2.46) \times 10^{-14}$ \\

\hline
\hline
\label{tab:meanave}
\end{tabular}
\end{center}

\vspace{0.8cm}
\end{table}

\begin{figure}
\includegraphics[width=0.45\textwidth]{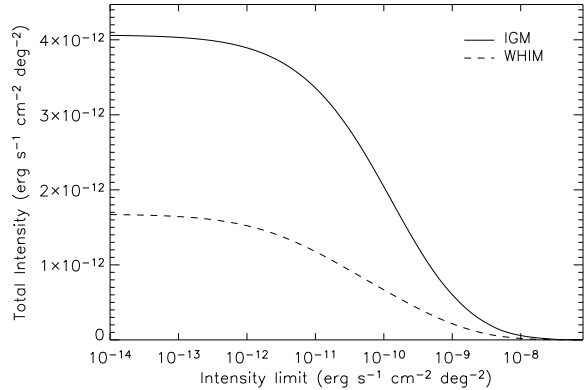}
\caption{
The value of the mean intensity (averaged over ten different map
realizations) as a function of the lower limit of the pixel surface
brightness. The solid and dashed lines refer to the IGM and the WHIM,
respectively.}
\label{fig:limit}
\end{figure}

To determine which sources contribute most to the soft 
X-ray emission, in Fig. \ref{fig:limit} we show how the mean intensity
varies when we include pixels with fluxes above a given threshold.  The
results show that the bremsstrahlung emission is dominated by very
bright objects: pixels having a surface brightness larger than
$10^{-10}$ \sbunits \  (mainly corresponding to galaxy clusters and
groups as shown in Section \ref{sect:separating}) contribute to more than 50 
per cent of the total value, for both the IGM and the WHIM.

\begin{figure*}
\includegraphics[width=0.32\textwidth]{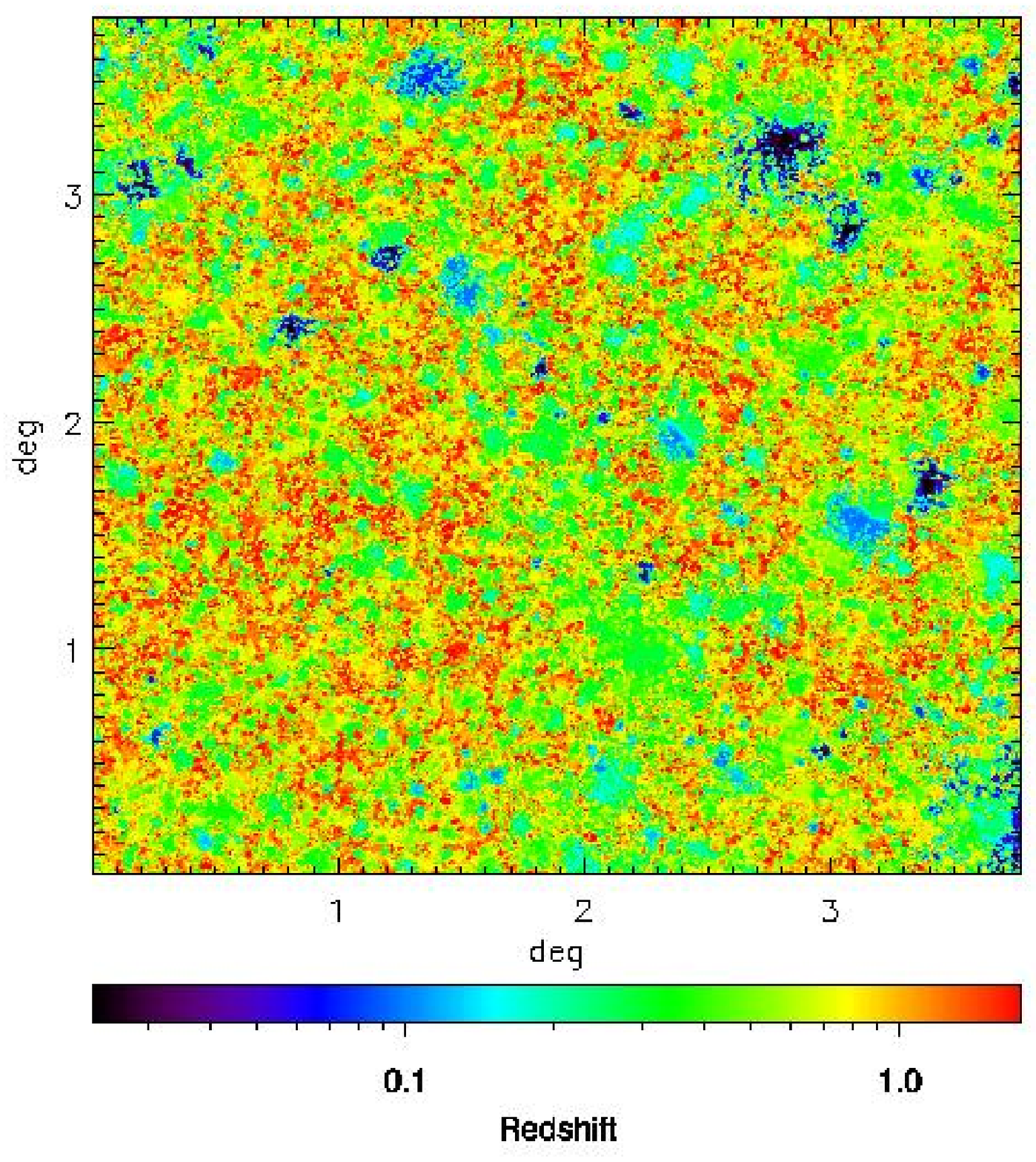}
\includegraphics[width=0.32\textwidth]{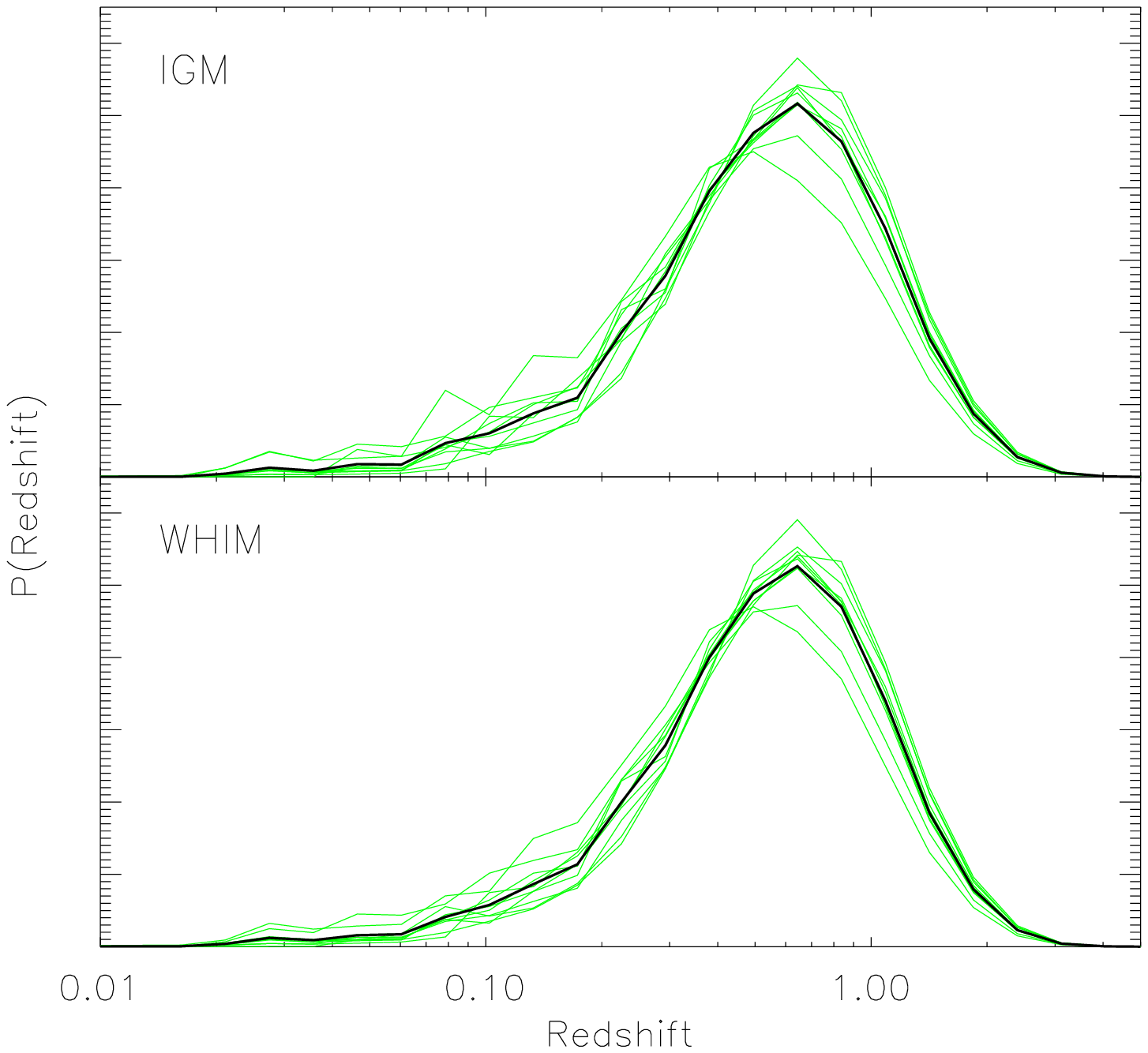}
\includegraphics[width=0.32\textwidth]{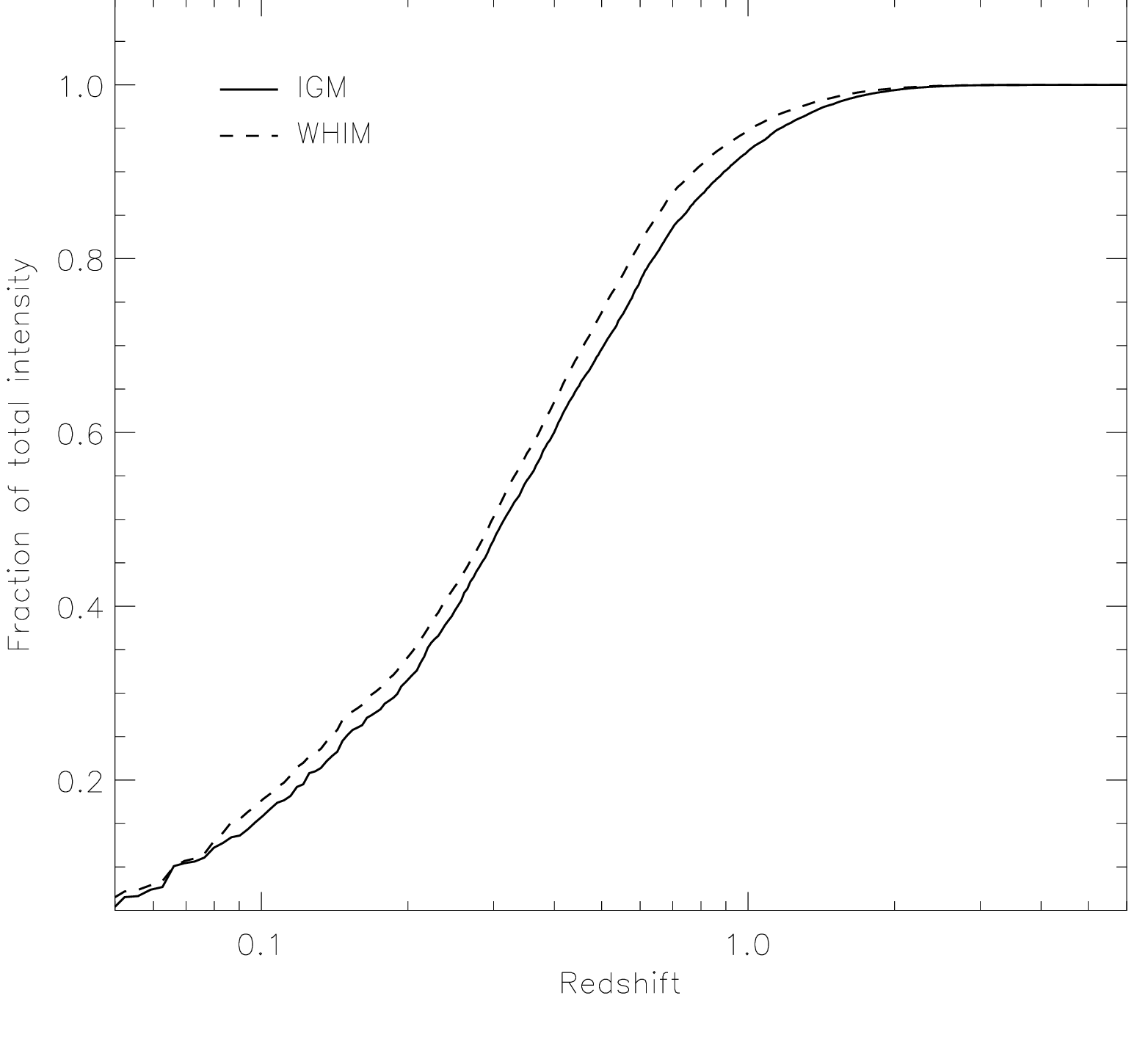}
\caption{
Left panel: the map of the flux-weighted redshift, computed for the
IGM in the soft X-ray band.  The plot refers to the same light-cone
realization shown in Fig. \ref{fig:softmaps}. Central panel: as in Fig.
\ref{fig:softdistr}, but for the distribution of the flux-weighted 
redshift in the soft X-ray band.  Right panel: the integrated soft
X-ray intensity (normalized to the total value) as a function of
redshift. The solid and dashed lines refer to the IGM and the WHIM,
respectively. The values plotted are the mean of ten map
realizations. }
\label{fig:z_soft}
\end{figure*}

In order to have an indication of the typical distances of the soft
X-ray sources, we follow \cite{croft2001} and compute, for each pixel,
the mean redshift of the particles contributing to the XRB weighted by
their fluxes. The resulting map for the same IGM realization shown in
the left panel of Fig. \ref{fig:softmaps} is displayed in
Fig. \ref{fig:z_soft} (left panel). Almost all the bright clusters are
at low redshift ($z<0.15$), while the emission from the faintest
structures mainly comes from redshifts around unity.  This is
confirmed by the distributions of the flux-weighted redshifts (central
panel of Fig. \ref{fig:z_soft}), computed from the complete set of ten
realizations: the shapes of the distributions for the IGM and the WHIM 
are very similar,
with a peak at $z\approx 0.7$.  Notice that few realizations have an
excess of contribution from sources at relatively low redshift
($z<0.15$): this is the origin of the large spread of 
the mean intensity previously discussed.  The analysis of the maps
also clearly shows how the projection of nearby and distant objects
makes it very difficult to see imprints of the filamentary network
present in the large-scale structure of the universe.  Finally, in the
right panel of Fig. \ref{fig:z_soft} we show the integrated soft X-ray
intensity as a function of the limiting redshift. The curves for both 
the IGM and the WHIM reach half the total value at $z\simeq 0.3$,
the 90 per cent level is reached around $z\simeq0.9$ for the IGM and
at $z\simeq0.8$ for the WHIM. We can also notice that the sources at
$z>2$ contribute less than 1 per cent of the soft X-ray flux.

\subsection{The autocorrelation function}

We analyze the clustering properties of the IGM in the soft X-ray band 
by estimating the autocorrelation function in our simulated maps.
Following \cite{croft2001}, we calculate the contrast
$\delta(\vec{x})$ defined as
\begin{equation}
\delta(\vec{x}) \equiv [I_X(\vec{x})/\overline{\cal I}]-1 \ ,
\end{equation}
where $I_X(\vec{x})$ is the surface brightness at the
position $\vec{x}$ and $\overline{\cal I}$ is the average of each
map. The contrast is then used to compute the angular correlation function
as
\begin{equation}
w(\theta)=\langle \delta(\vec{x}) \delta(\vec{x}+\vec{\theta})\rangle \ .
\end{equation}

\begin{figure}
\includegraphics[width=0.45\textwidth]{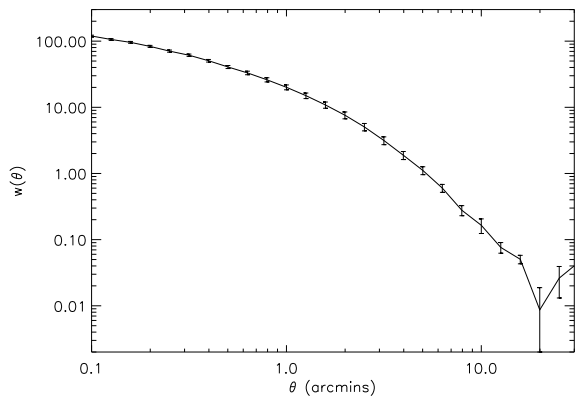}
\caption{
The angular correlation function of the IGM in the soft (0.5-2 keV) X-ray
band.  The solid line is the average over ten
different maps 3.78 deg on a side; error bars are the error on the mean
calculated from the scatter of the maps.}
\label{fig:acf_soft}
\end{figure}

The results are shown in Fig. \ref{fig:acf_soft}, where we plot the
average of the angular correlation functions calculated by considering
the ten different realizations of the light cone; error bars are the 
error on the mean obtained from the scatter of the maps. We notice that the
mean function is always positive up to $\theta\sim30^\prime$, although
some of the realizations show negative values for $\theta \ga
8^\prime$; the smallest scale at which $w(\theta)$ is compatible with
a null value is $20^\prime$. From the plot it is also evident a slow
change of the logarithmic slope. This is confirmed by our analysis: by
fitting our mean function under the assumption of a power-law
relation, $w(\theta)\propto \theta^\alpha$, we find $\alpha\approx
-0.9$ and $\alpha\approx -2.6$ in the intervals $0.1^\prime < \theta <
2^\prime$ and $2^\prime < \theta < 30^\prime$, respectively.

Our results can be directly compared to those obtained by
\cite{croft2001}.  In general the  autocorrelation function 
extracted from our maps is larger by a factor of about 6.  Moreover we
find that at large
scales the decrease of their $w(\theta)$ is slower (they found
$\alpha=-1.4$). These discrepancies can be easily explained
by considering the differences between the two simulations. First, and
most importantly, the size of the simulation box used by
\cite{croft2001} is about four times smaller than our box size: the
resulting lack of large-scale power can lead to a significant
underestimate of the clustering strength. Second, the physical
processes included in our simulation produce higher values of the
soft X-ray signal: as a consequence, the contrast $\delta$ also tends
to be amplified, thus originating a larger autocorrelation function.

\section{The contribution in the hard X-ray band}\label{sect:hardX}

\begin{figure*}
\includegraphics[width=0.45\textwidth]{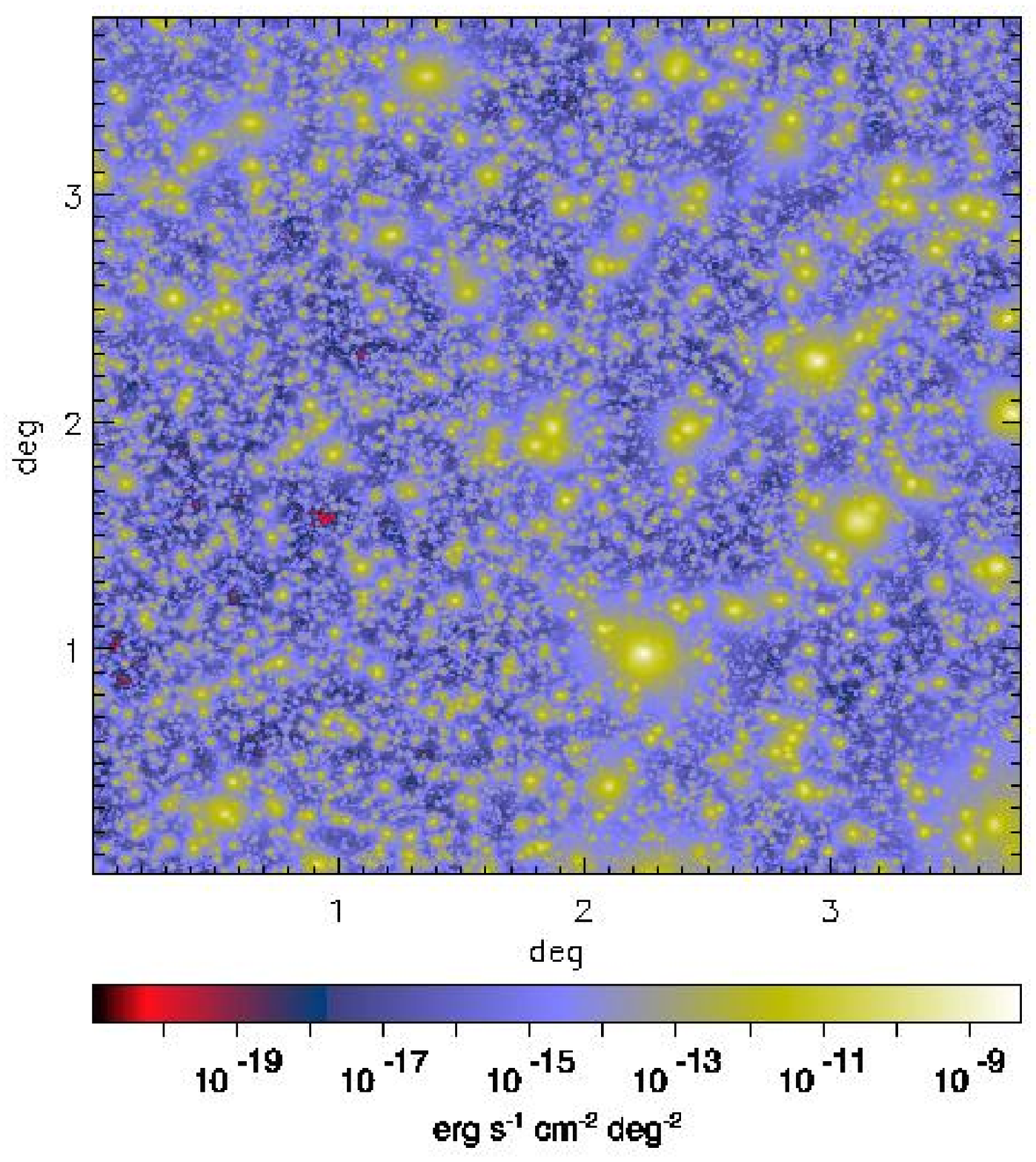}
\includegraphics[width=0.45\textwidth]{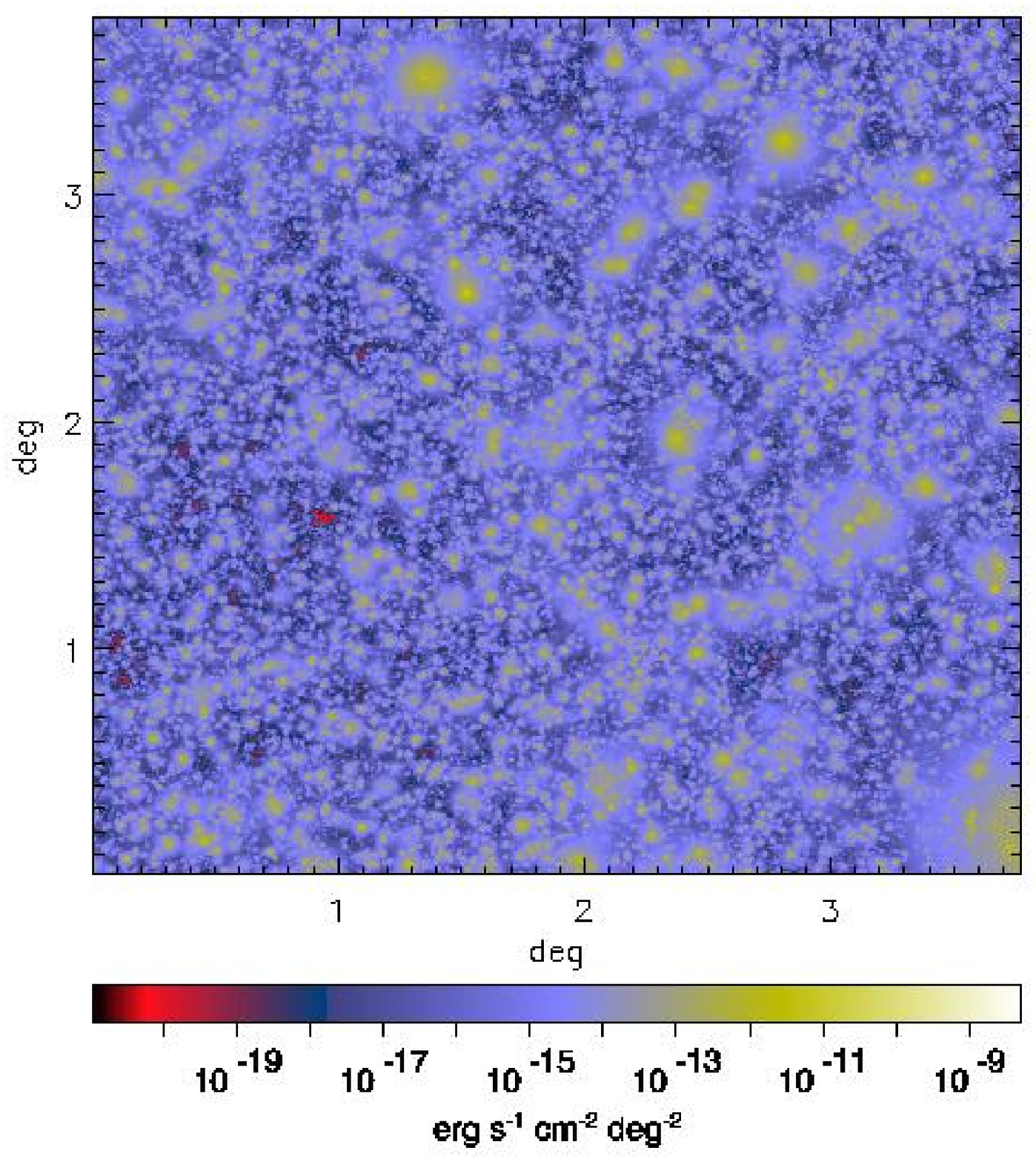}
\caption{
Maps of the hard (2-10 keV) X-ray intensity for the IGM and the WHIM (left
and right panels, respectively).  The map is 3.78 deg on a side
and the pixel is 1.66\arcsec on a side.  These maps have been
obtained from the same light-cone realization shown in
Fig. \ref{fig:softmaps}.}
\label{fig:hardmaps}
\end{figure*}

It is well known that the hard (2-10 keV) X-ray band is strongly
dominated by the emission from AGN. However, a small portion of gas,
mainly located within the richest clusters, is 
at high temperature and also gives
some contribution in this band. The maps presented in
Fig. \ref{fig:hardmaps} show the hard band intensity for the IGM and
the WHIM (left and right panels, respectively) for the same light-cone
realization shown in Fig. \ref{fig:softmaps}. We note that, even if
in some positions, corresponding to the hottest clusters, the
intensity can reach values similar to those obtained in the soft X-ray
maps ($10^{-9}$ \sbunits), outside virialized objects the hard X-ray
maps are much fainter. In the map for the WHIM the signal is strongly
reduced and limited to the cluster atmospheres, with no evident
emission from the diffuse gas: this makes the cosmic web not
observable in this band because of its very low signal-to-noise ratio and
projection effects. However, the high contrast between clusters and
diffuse regions in the hard band maps suggests an empirical way to
identify the positions of galaxy clusters in our simulated maps. This
method will be used in Section \ref{sect:diffusegas} to separate the emission
coming from extended objects from that produced by diffuse gas.

The small contribution of the IGM and the WHIM to the hard XRB is
confirmed by the average total intensity reported in Table
\ref{tab:meanave}, for both components: we find  about $1 \times 10^{-12}$ 
\sbunits \ and $3 \times 10^{-14}$ \sbunits, for the IGM and the WHIM, 
respectively.  These values account for approximately 25 per cent and
2 per cent of the corresponding mean intensity in the soft band. 
Note that, while for the WHIM our flux is a factor of
two larger than \cite{croft2001} result, for the IGM there
is a substantial agreement. In
fact, the lack of large-scale power in the simulation by 
\cite{croft2001}, due to the small box size, 
is compensated by the higher dynamical evolution associated
to 
the higher value of $\Omega_{\rm m}$, for a fixed power spectrum
normalization $\sigma_8$ assumed in both analyses. Moreover, the effect of feedback 
on the hard
X-ray emission by rich clusters is expected to be less relevant.
We have to consider, however, that the field-to-field variance is
quite large for the hard X-ray band, because most of the signal comes
from nearby bright clusters: in fact, the hard X-ray spectrum decreases
exponentially and, even for hot objects, the bulk of the
signal is rapidly shifted into the soft band at increasing redshifts.

\begin{figure*}
\includegraphics[width=0.45\textwidth]{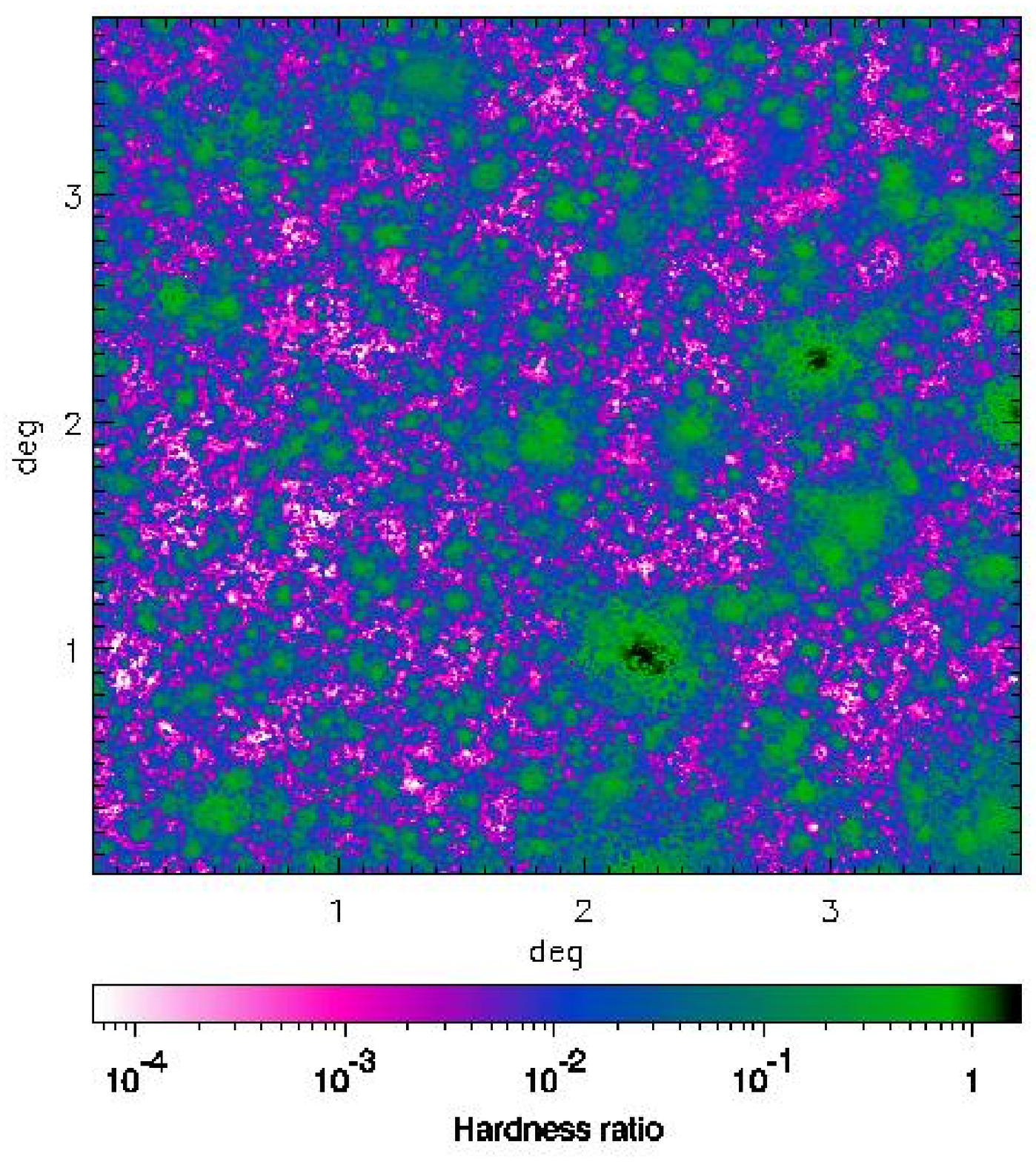}
\includegraphics[width=0.45\textwidth]{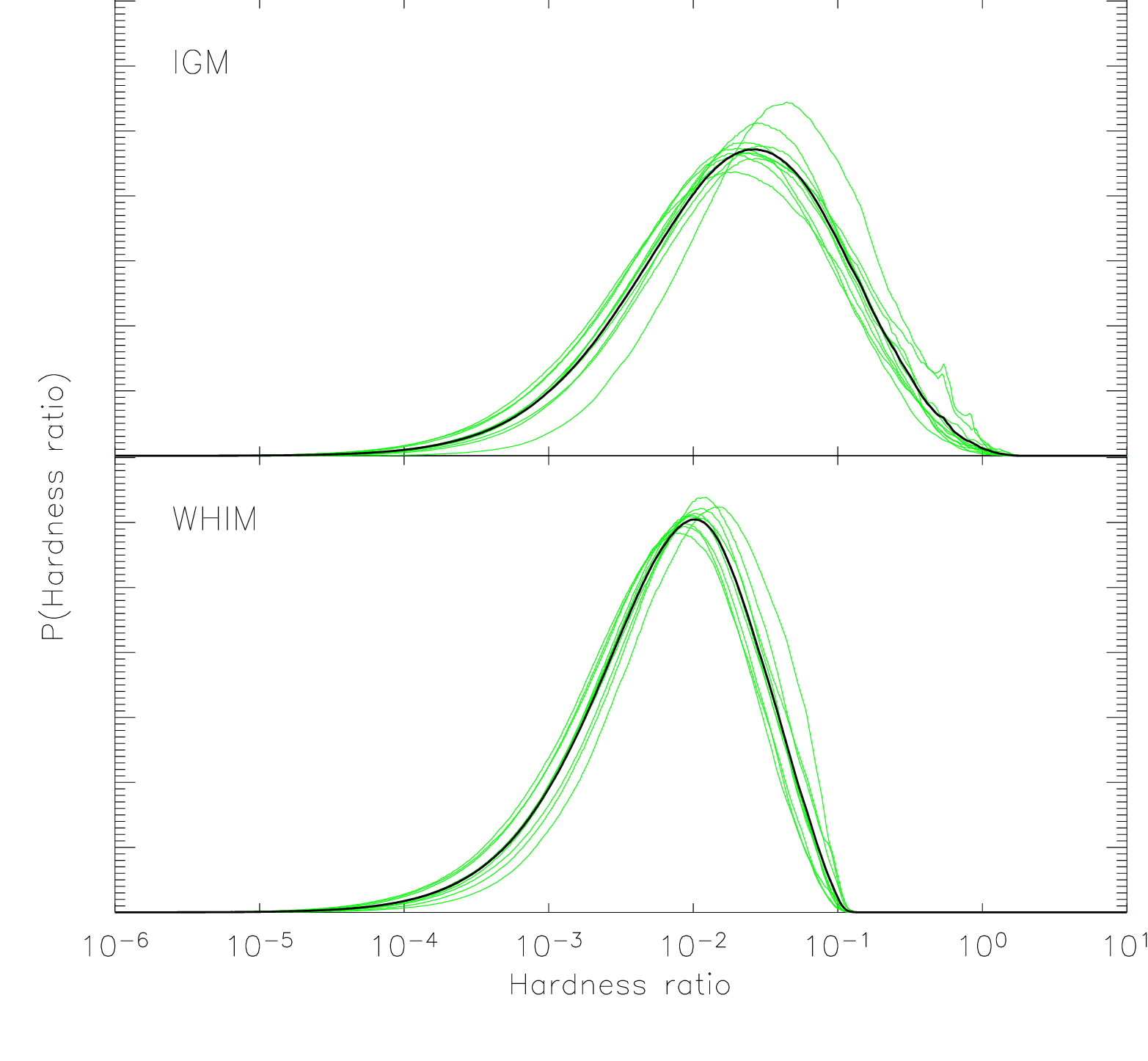}
\caption{
Left panel: the map of the hardness ratio (hard band to soft band)
$HR$.  This map is obtained from the same realization of the IGM
light-cone shown in the left panels of Figs. \ref{fig:softmaps} and
\ref{fig:hardmaps}.  Right panels: distribution of $HR$ for the IGM
(up) and for the WHIM (down). The thin lines show the results for ten
different realizations; the thick solid line is the corresponding average.
}
\label{fig:ratio}
\end{figure*}

A way to discriminate the AGN signal from that coming from clusters
and the cosmic web is based on the energy flux hardness ratio ($HR$), here
defined as the ratio between the intensities in the hard and soft
bands at each pixel position: $HR\equiv I_X[2-10 {\rm keV}]/I_X[0.5-2
{\rm keV}]$. The emission from AGN is in fact expected to have
$HR>0.5$, while much lower values are common for galaxy clusters and
diffuse gas. The map for the $HR$ obtained from the same IGM
realization displayed in Figs. \ref{fig:softmaps} and \ref{fig:hardmaps}
is shown in the left panel of Fig. \ref{fig:ratio}: in the plot it is
easy to recognize the position of the richest galaxy clusters, while
the connecting filamentary structure is almost completely absent.  The
distributions of the $HR$ values for the ten realizations and their
average are shown in the right panels, for both the IGM (up) and 
the WHIM (down). In the former case the median value is around
$HR=0.03$, and less than a few per cent of the pixels have $HR>0.5$; in the
latter case the median is smaller than 0.01 and no pixel has $HR>0.1$.
  
\section{Comparison with previous works} \label{sect:comparison}

In this paper we study the properties of the X-ray background from the
large-scale structure of the universe. In particular we followed the
same method to build light-cone realizations applied by
\cite{croft2001}, performing then similar statistical tests.
 From a comparison between our and their analysis, we found that in our
soft X-ray maps
the mean intensities of the IGM and the WHIM are larger than the intensities
obtained by \cite{croft2001} by a factor of 1.8 and 4, respectively.
The autocorrelation function of the signal is also larger by a factor
of about 6. As for the hard X--ray band, we find a substantial agreement
with \cite{croft2001} for the
IGM contribution, while our contribution from the WHIM is a factor of 2
higher.

These differences can be understood in terms of the different
characteristics of the two hydrodynamic simulations on which the
analyses are based. The main reason for our higher values is that we
followed the evolution of a much larger (almost 60 times) volume. This
allows us to obtain a better representation of very massive clusters
that contribute to most of the emission (see the discussion in Section
\ref{sect:softX}). The larger sampled volume is also the main motivation
of our higher autocorrelation function, which includes the contribution
of longer wavelength modes.

However, an important difference between the two hydrodynamic
simulations is also related to the treatment of the physical processes
affecting the evolution of the baryonic component. The simulation of
\cite{croft2001} did not consider the possible multiphase nature of
the starforming gas. For this reason they had to resort {\it a
posteriori} to a correction of their results to account for it.  On
the contrary, the GADGET-2 code used for our simulation follows the
different physical processes (radiative cooling, star formation and
supernova feedback) in a self-consistent way. In particular the
inclusion of a phenomenological description of galactic winds provides
a much more efficient energy feedback than that implemented in the
simulation by \cite{croft2001}. Moreover the code directly considers
the effect of a photoionizing time-dependent UV background. Globally
this gas treatment produces a larger fraction of WHIM. This explains
why larger differences are found for the contributions of the WHIM,
while less significant differences are found between the signals
in the hard X-ray band, which are mostly originated by the
hottest objects.  We also assumed a cosmological model which is closer
to that suggested by the most recent observational data with
$\Omega_{\rm m}=0.3$, while \citealt{croft2001} adopted $\Omega_{\rm
m}=0.4$: thus we have a higher contribution from baryons at high
redshifts due to the slower dynamical evolution of the structures.
The apparent agreement of our result for the hard band emission of the
IGM with \cite{croft2001} is due to a combination of all of these
factors (see the discussion in Section \ref{sect:hardX}).

Finally, a similar result to the one of \cite{croft2001} has been
obtained also by \cite{bryan2001} for the soft X-ray emission of the IGM
from their AMR simulation which includes a simple feedback model. Again
the difference with respect to the present work is maninly due to the
relatively small volume sampled by their simulation.


\section{The contribution
of diffuse gas to the soft X-ray background}\label{sect:diffusegas}

The goal of this section is to extract, from the soft X-ray maps
discussed above, the contribution produced by the cosmic web,
i.e. the baryons which are not included in the groups and clusters of galaxies 
currently identified in deep X-ray surveys. In order to
check whether the modelling of the physical processes treated by
our hydrodynamical simulation (which also influence the low-density
gas regions) is reliable, we compare our results with
observational estimates (mainly upper limits).  We have to notice that
in the literature, as reviewed in the following subsection, there is
no robust determination of the contribution of diffuse gas to the soft XRB: 
the value that we will use later for comparison is based, to
some extent, on interpretation of the existing data.
 
\subsection{Observational estimates}
\label{sect:obs_est}

In recent years, thanks to the availability of new X-ray
satellites like \emph{Chandra} and \emph{XMM/Newton}, there have been
many efforts to measure the total XRB and to resolve it
in its different components \citep[see, e.g.][ and references therein]
{giacconi2002,moretti2002,alexander2003,worsley2004,bauer2004}.  In
fact the possibility of recognizing the discrete sources which 
give the main contribution opens new windows to study the connection
between AGN and galaxy formation and, most importantly for this work,
constrains the possible signals from the diffuse gas.

At present, the most reliable estimate of the total XRB in the soft
(0.5-2 keV) band has been obtained by \cite{worsley2005}.  By assuming
the shape of the XRB in the [1-8 keV] band determined by \cite{deluca2004}
from \emph{BeppoSAX} data, and by taking into account the steepening of the
spectrum below 1 keV as found by \cite{roberts2001}, \cite{worsley2005} estimate an
${\rm XRB}[0.5-2 {\rm keV}] = (8.12\pm 0.23) \times 10^{-12}$
\sbunits.  Note that this value is larger than the one adopted by
\cite{bauer2004}, who found ${\rm XRB}[0.5-2 {\rm keV}] = (7.52\pm
0.35) \times 10^{-12}$\sbunits. The discrepancy is originated by the
fact that \cite{bauer2004} extrapolate the hard spectrum in the soft
band by assuming a slope of $-1.4$: of course, this does not take into
account the steepening due to the non-AGN contribution, which, by the
way, is the one we are most interested in.  This total value includes
any contribution, namely the Local Hot Bubble, the Galactic halo,
unresolved galactic stars, and most importantly, the extragalactic
components (both bright and faint) from point sources (AGN), clusters,
groups and the truly diffuse gas.  The resolved (to date) XRB in
the soft band produced by these extragalactic sources has been
recently estimated by \cite{worsley2005}, who found $(6.9\pm 0.2)
\times 10^{-12}$ \sbunits. Similar results for the resolved component
are presented by \cite{bauer2004}: $(6.7\pm 0.3) \times 10^{-12}$
\sbunits.  Therefore, by assuming the values found by \cite{worsley2005}
for the total XRB in the soft band, the emission which is still
unresolved is approximately $(1.2\pm 0.3) \times 10^{-12}$
\sbunits. This value represents an upper limit to the possible
contribution from the diffuse gas, because it still includes the unknown
Galactic and Local Hot Bubble contributions. Notice that the latter components
probably dominate the XRB below 0.5 keV, making highly uncertain
the estimate of the extragalactic contribution in the lowest X-ray
energy band, where the signal from the cosmic web is expected to be
even more significant.

\subsection{Separating the diffuse soft X-ray emission from
galaxy groups and clusters in the simulated maps}  \label{sect:separating}

By comparing naively the estimates of the IGM soft X-ray
contribution reported in Table \ref{tab:meanave} with the observational
data discussed above, we would notice that our simulated XRB accounts for about 50 
per cent
of the observed value of the XRB, but it is much larger (by a factor 
4) than the observed upper limit obtained by removing the discrete sources.
Even considering the signal from the WHIM only, i.e. from gas having
temperature between $10^{5}$ and $10^{7}$ K, the signal in our maps would be
too large.  However, this comparison is quite misleading, because in
the results from the simulation we still include the portion of the warm
gas located in groups or in the external atmospheres of galaxy
clusters; on the contrary, this gas is not present in the observed
value $1.2\pm 0.3 \times 10^{-12}$ \sbunits.  For this reason we
need a more detailed analysis of our maps, where these objects need to
be excluded.

We already know, from the previous analysis of our original
hydrodynamical simulation \citep[see, e.g.,][]{borgani2004}, that the
modeling of the included physical processes produces overluminous
groups and poor clusters of galaxies. As discussed in that paper,
additional sources of energy (like feedback from SN-Ia or AGN) would
help to solve this discrepancy. This excess is located in
high-density regions, whereas we are interested in the soft X-ray
emission from the cosmic web: therefore, we need to identify and
remove the X-ray extended sources, which correspond to galaxy clusters
and groups, from the total intensity of our maps.

The proper procedure would be to identify and remove all the clusters
and groups from mock observations of our simulated volumes.  In
practice, this is unfeasible, because it would require simulating a
full wide--area survey with all the related observational aspects,
which is far beyond the aim of this paper.  Therefore, we will remove
clusters and groups on the basis of their total flux as it appears on
our flux maps. To establish a definition of a cluster or a group to be
applied to our bidimensional maps, we use an identification criterion
based on the surface brightness value.  We identify as an X-ray
extended source (group or cluster) all the connected regions of pixel
above a surface brightness threshold. 

To accomplish this, we use the maps in the hard (2-10 keV) band
because the emission from the diffuse gas in this band is much lower than in
the soft band and therefore it is easier to identify the regions corresponding to 
virialized haloes
(see Fig. \ref{fig:hardmaps}).  Then, we choose the value of the
surface brightness threshold by inspecting the faintest groups
identified in the deepest X-ray observations to date.  Specifically,
we consider a group (CDFS-594) identified in the CDF South \citep{giacconi2002}
at $z \sim 0.7-0.8$.  Its emission is detected with a high
signal-to-noise ratio over a circular region of approximately 2700
arcsec$^2$.  After fitting its spectrum with a \emph{mekal} model
\citep[see, e.g.,][ and references therein]{liedahl1995} we obtain a
temperature of about 2.1 keV.  Therefore, the corresponding
\emph{average} surface brightness in the hard band is $3.06 \times
10^{-12}$ \sbunits, which will be used as the reference value in the
following analysis\footnote{Similar results are obtained by
considering a different galaxy group (CDFS-645), again identified in
the CDF South exposure.}.  Notice that this average value corresponds
to the central part of the group (typically about 3 core radii), which
is the region where the emission can be realistically detected.
However, some emission associated to the X-ray halo is expected to
provide sub--threshold contribution up to the virial radius. Since we
want to remove all the X-ray emission associated to a given halo, in
the following analysis we will also consider smaller, up to 8 times, surface brightness
thresholds.  We choose these thresholds, because the surface brightness
of an isothermal gas in the central region (about 3 core radii)
is roughly 8 times larger than at the virial radius.
Finally, we consider a threshold for the minimum angular size of the
region that can be identified as a group or a cluster.  For this
purpose we choose a reference physical length of 100 kpc, that
corresponds to a minimum angular size of $\sim$10 arcsec (at $z\sim
1$) for the cosmological model assumed in our simulation: this gives a
minimum angular surface of $\sim$310 arcsec$^2$.  This quantity
roughly corresponds to the minimum size of the central detectable regions
of X-ray groups (see \citealt{willis2005}).  This last criterion has a
small impact on the final results (i.e. the expected average surface brightness 
of the cosmic web changes of few percent when doubling the minimum angular 
size of the haloes).

\subsection{The soft X-ray emission from the cosmic web in the 
simulated maps}

By applying the previously discussed thresholds of the surface brightness
and size on the maps produced in the hard (2-10 keV) X-ray band, we
create a catalogue of connected regions corresponding to the extended
objects like groups and clusters. We use them as a mask on the soft
(0.5-2 keV) X-ray maps to remove their emission and finally obtain the
diffuse contribution.  By averaging over our ten different realizations,
we find that the signal from the diffuse gas is 
$1.57 \times 10^{-12}$ \sbunits\, which is comparable to the
observed upper limit. 69 per cent of the diffuse emission 
comes from gas with temperature in the range $10^5-10^7$ K, whereas the
total contribution of clusters and groups to the XRB is $2.49 \times
10^{-12}$ \sbunits; this value is larger than the total contribution from
clusters and groups estimated from the observed number counts
\citep[][ see also the corresponding discussion at the end of this
section]{rosati2002}.  As noted above, the adopted surface
brightness threshold can be considered as an average of the group
emission at intermediate redshifts, and then it could be too high to
account for all clusters' and groups' emission.  In fact we find that
31 per cent of the $1.57 \times 10^{-12}$ \sbunits \ emission is still
produced by gas having a temperature larger than $10^7$ K.  This is
further confirmed by a visual inspection of the masked maps: the
external regions of nearby clusters are not excluded. This is an
indication that in general the surface brightness threshold defining
high density regions associated to groups and clusters of galaxies is
probably lower than the adopted value. For this reason, we repeat the
previous analysis by using thresholds which are smaller by factor of
2, 4 and 8.  The results, reported in Table \ref{tab:sbcut}, show a
reduction of the contribution of the diffuse gas by a factor of 2 when the
surface brightness cut is 8 times smaller; the corresponding fraction
of the signal from the WHIM increases up to 82 per cent. In general we can
conclude that the soft X-ray emission from the diffuse gas computed from
our simulation is consistent with the upper limit coming from
observational data.  The feedback model we are assuming is 
efficient in the WHIM and/or in regions with intermediate density, although
it can show problems in the deepest potential wells.

\begin{table*}
\begin{center}
\caption{
Contribution to the soft (0.5-2 keV) XRB of galaxy clusters and groups
(Column 2) and  of the diffuse gas (Column 3) as a function of the adopted
threshold of the surface brightness (Column 1).  We also report the
percentage of the diffuse contribution coming from the WHIM (Column
4). The quoted errors are the r.m.s. in fields of 1 deg$^2$.  All
values are in units of $10^{-12}$ \sbunits.}
\begin{tabular}{cccc}
\hline
\hline
\multicolumn{4}{c}{\scshape{Soft (0.5-2 keV) X-ray background}}\\

\multicolumn{4}{c}{($10^{-12}$ \sbunits)} \\
{\scshape{surface brightness threshold  }} 
 &  \multicolumn{3}{c}{\scshape{Contributions from}}\\
\scshape{in the hard (2-10 keV) X-ray band} &  \scshape{clusters and groups} 
& \scshape{Diffuse gas} & \%WHIM \\
\hline

$3.06$ & $2.49\pm2.04$ & $1.57\pm0.27$ & 
($69\pm13$)\% \\
$1.53$ & $2.78\pm2.10$ & $1.28\pm0.19$ & 
($73\pm12$)\% \\
$0.77$ & $3.05\pm2.14$ & $1.01\pm0.13$ & 
($78\pm11$)\% \\
$0.38$ & $3.28\pm2.17$ & $0.78\pm0.09$ & 
($82\pm11$)\% \\

\hline
\scshape{observational upper limit}     & & $1.2\pm 0.3$\\
\hline
\hline
\label{tab:sbcut}

\end{tabular}
\end{center}
\end{table*}

\begin{figure}
\includegraphics[width=0.45\textwidth]{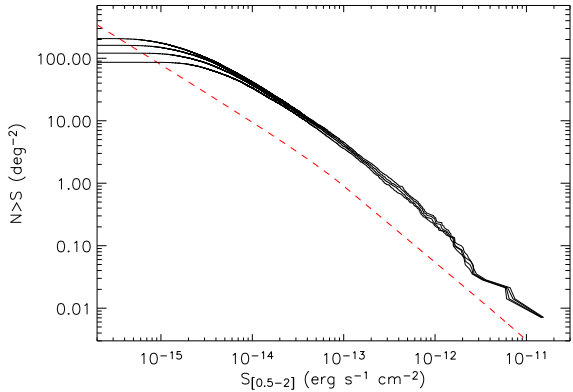}
\caption{
Number of objects as a function of the limiting flux in the soft
(0.5-2 keV) X-ray band.  The black solid lines are the results
obtained by considering the connected regions with four different surface
brightness cuts (see Table \ref{tab:sbcut} for the adopted values).
For comparison, we show (red dashed line) the observational data
obtained by \protect \cite{rosati2002} by combining the results from
the \emph{Rosat} Deep Cluster Survey \protect \citep{rosati1998}, EMSS
\protect \citep{rosati1995}, BCS \protect \citep{ebeling1998} and
REFLEX \citep{bohringer2001}. }
\label{fig:lognlogs}
\end{figure}

Finally, as a self-consistency check, we compute the \lognlogs\
function (always in the soft X-ray band) for our connected
regions. The results are shown in Fig. \ref{fig:lognlogs} for the four
different thresholds reported in Table \ref{tab:sbcut}.  Again, we
notice the excess with respect to the observational data, represented
by a combination of data made by \cite{rosati2002} and coming from
the \emph{Rosat} Deep Cluster Survey \citep{rosati1998}, EMSS
\citep{rosati1995}, BCS \citep{ebeling1998} and REFLEX
\citep{bohringer2001}. This excess is produced by the
overluminous objects which are present in our simulation at all
relevant redshifts: for this reason, number counts are overestimated over
the entire flux range. Finally, the apparent flattening at low fluxes
($S\approx 10^{-15}$\lumunits) is originated by both resolution
effects and the surface brightness limit assumed for the
identification of the extended sources.

\section{Conclusions}\label{sect:conclusions}

In this paper we have used a cosmological hydrodynamical simulation of
a concordance $\Lambda$CDM model to discuss the properties of the
diffuse XRB.  The simulation \citep{borgani2004} includes 
many relevant physical processes affecting the gas
component. Indeed the numerical treatment accounts for a
time-dependent, photoionizing UV uniform background, radiative cooling
processes within an optically thin gas of hydrogen and helium in
collisional ionization equilibrium, star formation events, and feedback
processes from both SN-II and galactic winds. Our previous analyses
\citep{borgani2004,ettori2004} showed that the results of this
simulation are consistent with several observational 
X-ray properties of galaxy clusters. However, there is also a number of
discrepancies that remain unaccounted for: in particular the cluster
X-ray luminosity-temperature relation extracted from our simulation
appears too high, i.e. there are overluminous groups and small
clusters.

As suggested by different authors \citep[see,
e.g.,][]{voit2001a,bryan2001,xue2003}, an alternative way to
constrain the model describing the thermal properties of baryons
and their cosmic history is related to the soft X-ray emission from
diffuse gas.  Indeed, a model of heating and cooling of the
intergalactic medium makes predictions in terms of diffuse emission
from filaments and unresolved structures. These predictions can be compared with
the observational measurements of the XRB, which are now available (in
form of upper limits) thanks to the deep \emph{Chandra} data.

In order to test the model included in our simulation, we 
follow the method of \cite{croft2001} and, starting from
the simulation snapshots, we constructed a set of ten different
two-dimensional maps, of size $(3.78 {\rm deg})^2$, of the 
past light-cone back to $z=6$. The present analysis extends 
\cite{croft2001} previous work, thanks to a much larger volume sampling, to a
more realistic representation of the physical processes affecting the
baryon history and to the comparison with the observed XRB data after
the \emph{Chandra} era.

The main results obtained by using our set of maps are as follows:
\begin{itemize}
\item
The mean intensity of the IGM in the soft (0.5-2 keV) X-ray band is
about $4.1 \times 10^{-12}$ \sbunits; when considering the WHIM
(defined as gas with temperature between $10^{5}$ and $10^{7}$ K) the
mean intensity reduces to about $1.7 \times 10^{-12}$ \sbunits. The
distribution of the intensity in the maps is very similar to a
lognormal. 90 per cent of the signal comes from structures at
$z\lesssim0.9$.
\item
As expected, in the hard (2-10 keV) X-ray band the total mean
intensity is smaller (by a factor 4) than in the soft one and becomes
almost negligible when considering the WHIM (about $3 \times 10^{-14}$
\sbunits). The hardness ratio in the maps has a distribution which peaks at  
low values, with a median close to 0.03, which enables an easy
discrimination of the AGN signal.
\item
We obtain an estimate of $(0.8 - 1.6) \times 10^{-12}$ 
\sbunits \ for the soft X-ray emission 
from the diffuse gas after removing the regions corresponding to
extended objects, like galaxy groups and clusters, from the maps. Our result is 
consistent with the present upper limit coming from observational
data: $(1.2\pm 0.3) \times 10^{-12}$ \sbunits. This value,
recently obtained by \cite{worsley2005}, has been
estimated by removing the contribution of resolved sources (AGN,
groups and clusters of galaxies) from the total XRB.
\end{itemize}

As a conclusion, our results show that the physical processes 
discussed here are consistent with existing constraints on the 
X-ray properties of the warm-hot
baryons.  However, if in the near future the measurements of the
possible contribution from Galactic and local structures (still
included in the observational upper limits) will reduce 
the maximum allowed emission from the diffuse gas by a factor of
2 or more, our predictions
will start to conflict with the data.  In this case it will be required a more efficient
mechanism of feedback acting on the gas with intermediate
temperatures and densities: in fact the presence of a 
stronger feedback would place the gas on a higher adiabat, thereby 
preventing it from reaching high densities and, on turn, 
suppressing the diffuse emission.

Our analysis confirms that the comparison between observational
constraints on the diffuse emission in the soft X-ray band and results
from cosmological hydrodynamical simulations enables us to gain
information on the thermodynamical history of the diffuse baryons.  
In the future it will be interesting to study some
detailed simulated observations of the diffuse gas to check its
detectability not only through bremsstrahlung emission, but also 
via emission and/or absorption of oxygen lines (O{\sc
vi}, O{\sc vii} and O{\sc viii}): real observations
will probably be provided by some dedicated
wide-field high-resolution spectroscopic experiments now in
project. It will also be relevant to compare simulated 
results with observations performed by the next generation of
X-ray spectroscopic satellites, like \emph{Constellation-X/XEUS} or
\emph{NeXT}, provided that their field of view is large enough 
to cover the filamentary structure of the cosmic web.


\section*{acknowledgements}
Computations have been performed by using the IBM-SP4 at CINECA,
Bologna, with CPU time assigned under an INAF-CINECA grant. This work
has been partially supported by the PD-51 INFN grant. We wish to
thank the anonymous referee for useful comments that improved 
the presentation of our results.  We are grateful to
C. Gheller for his assistance. We acknowledge useful discussions with
S. Ettori, M. Galeazzi, P. Mazzotta, V. Springel, G. Tormen,
L. Tornatore and E. Ursino.

\label{lastpage}
\end{document}